\documentclass[
    ,final            
  ]
  {aipproc}

\layoutstyle{8x11double}

\input pix.sty

\usepackage{epsfig}
\usepackage{amsfonts}
\usepackage{axodraw}
\usepackage{amsmath}
\usepackage{bbm}

\newcommand{\sumint}[1]{\mbox{$\sum$}\!\!\!\!\!\!\!\int_{#1}}

\newcommand{\Nc}{N_{\rm c}}

\newcommand{\Tc}{T_{\rm c}}
\newcommand{\gB}{g_\rmii{B}}
\newcommand{\mE}{m_\rmii{E}}

\def\lsi{\raise0.3ex\hbox{$<$\kern-0.75em\raise-1.1ex\hbox{$\sim$}}}
\def\gsi{\raise0.3ex\hbox{$>$\kern-0.75em\raise-1.1ex\hbox{$\sim$}}}

\newcommand{\rmii}[1]{{\mbox{\tiny\rm{#1}}}}

\newcommand{\Tint}[1]{{\hbox{$\sum$}\!\!\!\!\!\!\!\int\,}_{\!\!\!\!\raise-0.9ex\hbox{$\scriptstyle{#1}$}}}
\newcommand{\Tinti}[1]{{{\Sigma}\!\!\!\!\raise0.3ex\hbox{$\int$}_\rmii{${#1}$}}}
\renewcommand{\Tint}[1]{\sumint{#1}}

\newcommand{\bi}{\begin{itemize}}
\newcommand{\ei}{\end{itemize}}

\newcommand{\hide}[1]{ }

\def\TAsc(#1,#2)(#3,#4,#5)%
{\SetWidth{2.0}\CArc(#1,#2)(#3,#4,#5)\SetWidth{1.0}}
\def\Lwidth{3}

\def\TAgl(#1,#2)(#3,#4,#5){\SetWidth{2.0}\PhotonArc(#1,#2)(#3,#4,#5){\Lwidth}%
{6.283 #3 mul 360 div #4 #5 sub #4 #5 sub mul sqrt mul Tdensity mul}%
\SetWidth{1.0}}
\def\TLgl(#1,#2)(#3,#4){\SetWidth{2.0}\Photon(#1,#2)(#3,#4){\Lwidth}
{#1 #3 sub #1 #3 sub mul #2 #4 sub #2 #4 sub mul add sqrt Tdensity mul}%
\SetWidth{1.0}}

\renewcommand{\picc}[1]{\;\parbox[c]{60pt}{\begin{picture}(60,30)(0,0)
\SetWidth{1.0}\SetScale{1.3} #1 \end{picture}}\; }
\def\Lwidth{1.3}

\newcommand{\picu}[1]{\;\parbox[c]{60pt}{\begin{picture}(60,30)(0,0)
\SetWidth{1.0}\SetScale{1.0} #1 \end{picture}}\; }
\def\EleA{\picu{%
 \Agl(30,5)(22.3,27,153)%
 \Agl(30,25)(22.3,207,333)%
 \COval(10,15)(2,2)(0){Black}{Black}%
 \COval(50,15)(2,2)(0){Black}{Black}%
}}
\def\EleB{\picu{%
 \Agl(30,5)(22.3,27,153)%
 \Agl(30,25)(22.3,207,333)%
 \COval(10,15)(2,2)(0){Black}{Black}%
 \COval(50,15)(2,2)(0){Black}{Black}%
 \Agl(58,15)(8,0,360)%
}}
\def\EleC{\picu{%
 \Agl(30,5)(22.3,27,153)%
 \Agl(30,25)(22.3,207,333)%
 \COval(10,15)(2,2)(0){Black}{Black}%
 \COval(50,15)(2,2)(0){Black}{Black}%
 \Lgl(10,15)(50,15)%
}}
\def\EleD{\picu{%
 \Agl(30,5)(22.3,90,153)%
 \Agl(30,25)(22.3,207,333)%
 \COval(10,15)(2,2)(0){Black}{Black}%
 \COval(50,15)(2,2)(0){Black}{Black}%
 \Agl(43,27)(12,180,300)%
 \Agl(38,16)(12,0,120)%
}}
\def\EleE{\picu{%
 \Agl(30,5)(22.3,27,153)%
 \Agl(30,25)(22.3,207,333)%
 \COval(10,15)(2,2)(0){Black}{Black}%
 \COval(50,15)(2,2)(0){Black}{Black}%
 \GCirc(30,27.3){4}{0.5}
}}
\def\EleF{\picu{%
 \Agl(20,10)(11.15,27,153)%
 \Agl(20,20)(11.15,207,333)%
 \Agl(40,10)(11.15,27,153)%
 \Agl(40,20)(11.15,207,333)%
 \COval(10,15)(2,2)(0){Black}{Black}%
 \COval(50,15)(2,2)(0){Black}{Black}%
}}
\def\EleG{\picu{%
 \Agl(30,5)(22.3,27,153)%
 \Agl(30,25)(22.3,207,333)%
 \COval(10,15)(2,2)(0){Black}{Black}%
 \COval(50,15)(2,2)(0){Black}{Black}%
 \Lgl(30,2.7)(30,27.3)%
}}
\begin{document}

\title{Scalar and Pseudoscalar Correlators in Yang-Mills Theory}


\author{Aleksi Vuorinen}{
  address={Faculty of Physics, University of Bielefeld, D-33501 Bielefeld, Germany}
}

\begin{abstract}
Correlation functions of the $FF$ and $F\tilde{F}$ operators in hot SU(3) Yang-Mills theory have recently been studied both on the lattice and in perturbation theory, and the results subsequently compared to the strong coupling limit of large-$N$ ${\mathcal N}=4$ Super Yang-Mills theory, available through the AdS/CFT correspondence. Here, we review the perturbative calculations, covering both Euclidean spatial correlators and the UV limit of spectral densities, and comment on the emerging physical picture as well as on possible extensions of the present studies.
\end{abstract}

\maketitle

\section{Introduction and Setup}

Studying the correlation functions of different gauge invariant local operators of QCD, or pure Yang-Mills theory, at high temperature is obviously of great interest from the point of view of the physics of the quark-gluon plasma and heavy quarks (see \textit{e.g.~}Refs.~\cite{hbm_d,Burnier:2007qm} and references therein). In Minkowski space, the infrared limit of the imaginary part of a retarded Green's function, \textit{i.e.~}the spectral density $\rho(\omega)$, contains information on the transport properties of the system, while in Euclidean coordinate space, the behavior of the correlators can be used to study other important physical phenomena, such as screening. Not surprisingly, these quantities have in recent years received considerable attention from both the lattice and perturbative sides; for an example regarding static observables, see Ref.~\cite{Iqbal:2009xz}. Lately, related studies in strongly coupled large-$N$ ${\mathcal N}=4$ Super Yang-Mills (SYM) theory have appeared as well, where the calculations have been performed using the gauge/gravity duality \cite{Iqbal:2009xz,kv}.

In these proceedings, we report on two recent perturbative studies of scalar (the trace of the energy momentum tensor, denoted below by $\theta$) and pseudoscalar (the topological charge density, $\chi$) correlators in pure SU(3) Yang-Mills theory \cite{lvv,Laine:2010fe}. The operators are defined through
\ba
 \theta \equiv c_\theta\, \gB^2 F^a_{\mu\nu}F^a_{\mu\nu}
 , \quad
 \chi \equiv c_\chi\, \epsilon_{\mu\nu\rho\sigma}
 \gB^2 F^a_{\mu\nu}F^a_{\rho\sigma}
 \;,
\ea
with $c_\theta \approx -\frac{b_0}{2}$, $c_\chi \equiv \frac{1}{64\pi^2}$ and $b_0 \equiv \frac{11\Nc}{3(4\pi)^2}$, while $\gB^2$ stands for the unrenormalized (bare) coupling constant.\footnote{Neither of these operators requires renormalization up to the two-loop order, in which we work.} The precise correlators under study are then
\ba
G_\theta(X) \equiv
\langle \theta(X) \theta(0) \rangle_\rmii{c}, \quad G_\chi(X) \equiv
 \langle \chi(X) \chi(0) \rangle ,
\ea
as well as the corresponding Fourier-transformed and temporally averaged versions thereof,
\ba
\bar{G}_\theta(x)\equiv \int_0^\beta {\rm d}\tau\, G_\theta(X)
 , \;
 \bar{G}_\chi(x)\equiv \int_0^\beta {\rm d}\tau\, G_\chi(X),\\
 \tilde G_\theta(P) \equiv \int_X e^{- i P\cdot X} G_\theta(X),\;\;\;\;\;\;\;\;\;\;\;\;\;\;\;\;\;\;\;\;\;\;\;\;\;\;\;\;\;\;\;\;\;\;\;\;\;\;\;\\
 \tilde G_\chi(P) \equiv \int_X e^{- i P\cdot X} G_\chi(X).\;\;\;\;\;\;\;\;\;\;\;\;\;\;\;\;\;\;\;\;\;\;\;\;\;\;\;\;\;\;\;\;\;\;\;\;\;\;\;
\ea
The method employed in these studies is perturbation theory evaluated up to and including next-to-leading order (NLO) corrections, and, as usual, infrared ambiguities are avoided by adding to the strictly perturbative result a contribution originating from the three-dimensional effective theory EQCD. For further technical details, we refer the reader to Refs.~\cite{lvv,Laine:2010fe}.

One of the main motivations for the work performed in Refs.~\cite{lvv,Laine:2010fe} was the somewhat surprising numerical result of Ref.~\cite{Iqbal:2009xz}, according to which the vacuum-subtracted correlators of the $\theta$ and $\chi$ channels show remarkably different behaviors at intermediate distances $r\sim 1/T$. This is at odds with expectations stemming from the corresponding calculations in free field theory and strongly coupled ${\mathcal N}=4$ SYM, both of which predict the two correlators to coincide. In addition to this, Ref.~\cite{Iqbal:2009xz} argues that the leading ultraviolet terms of the two correlators, dictated by the Operator Product Expansion (OPE), coincide, which, however, is not supported by the lattice data of the same reference.

\section{Methods}

%
\begin{figure}[t]
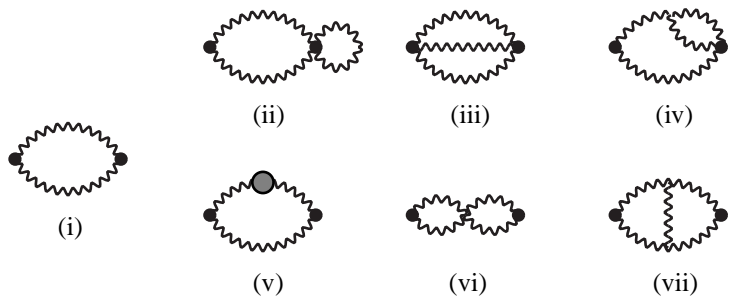


\hspace*{0.5cm}%
\begin{minipage}[c]{3cm}
\begin{eqnarray*}
&&
 \hspace*{-1cm}
 \EleA
\\[1mm]
&&
 \hspace*{0.0cm}
 \mbox{(i)}
\end{eqnarray*}
\end{minipage}%
\begin{minipage}[c]{8cm}
\begin{eqnarray*}
&&
 \hspace*{-1cm}
 \EleB \quad\;
 \EleC \quad\;
 \EleD \quad\;
\\[1mm]
&&
 \hspace*{0.0cm}
 \mbox{(ii)} \hspace*{2.2cm}
 \mbox{(iii)} \hspace*{2.2cm}
 \mbox{(iv)}
\\[5mm]
&&
 \hspace*{-1cm}
 \EleE \quad\;
 \EleF \quad\;
 \EleG \quad
\\[1mm]
&&
 \hspace*{0.0cm}
 \mbox{(v)} \hspace*{2.2cm}
 \mbox{(vi)} \hspace*{2.2cm}
 \mbox{(vii)}
 \\[3mm]
\end{eqnarray*}
\end{minipage}
  \caption{The Feynman graphs contributing to the $\theta$ and $\chi$ channel correlators up to two-loop order. The grey blob in (v) denotes the insertion of the one-loop gluon polarization tensor.}
\la{fig:graphs}
\end{figure}
%

A two-loop perturbative calculation of the correlation functions defined above obviously necessitates the evaluation of the two-point diagrams of Fig.~\ref{fig:graphs}. We begin the computation by first scalarizing the graphs in Euclidean momentum space, \textit{i.e.~}performing the Lorentz and color algebras in $\tilde G(P)$. To derive the leading terms of the UV expansions, considered in Ref.~\cite{lvv}, we then expand the 'master' sum-integrals in inverse powers of the external momenta, and subsequently take a Fourier transform to obtain the Euclidean coordinate space correlator $G(X)$ or the imaginary part of the Minkowskian momentum space expression to obtain the spectral density $\rho(\omega)$. For the intermediate distance spatial correlators of Ref.~\cite{Laine:2010fe}, we on the other hand specialize to the case of static external momenta and then proceed to write the sum-integrals in three-dimensional coordinate space, utilizing techniques developed in Refs.~\cite{az,Gynther:2007bw}. While most of the calculations are performed analytically, to obtain the final results requires evaluating some of the encountered integrals with numerical methods.

\section{Results}

\subsection{UV Asymptotics}

The main result of Ref.~\cite{lvv} is a two-loop expression for the UV expansions of the $\theta$ and $\chi$ channel correlators in Euclidean momentum and coordinate spaces as well as for the spectral densities in Minkowski space. The results thus obtained cover the full $T=0$ part of the correlators \cite{old} as well as the LO UV-terms of the finite-temperature parts, of which the latter are cast in the form of OPEs, \textit{i.e.~}as linear combinations of the unit operator, $e+p$ and $e-3p$ (see also Ref.~\cite{ope}). Perhaps the most important conclusion to be drawn from here concerns the difference between the two channels: Contrary to the claims of Ref.~\cite{Iqbal:2009xz}, we found that the coefficient of the $e-3p$ operator appears with opposite signs in the $\theta$ and $\chi$ channels, which at least partially explains the difference observed in the lattice results of the two correlators.

\subsection{Spatial Correlators at intermediate distances}

\begin{figure}
\caption{The behavior of the vacuum-subtracted parts of the $\theta$ (left) and $\chi$ (right) correlators, multiplied by $\bar{x}^3$, and compared with the corresponding OPE limits, at $T = 3 \Tc$. In both cases, the three lines correspond to three different choices of the renormalization scale $\bar{\Lambda}$, varied  by a factor of 2 around an 'optimal' scale.}
\includegraphics[height=.3\textheight]{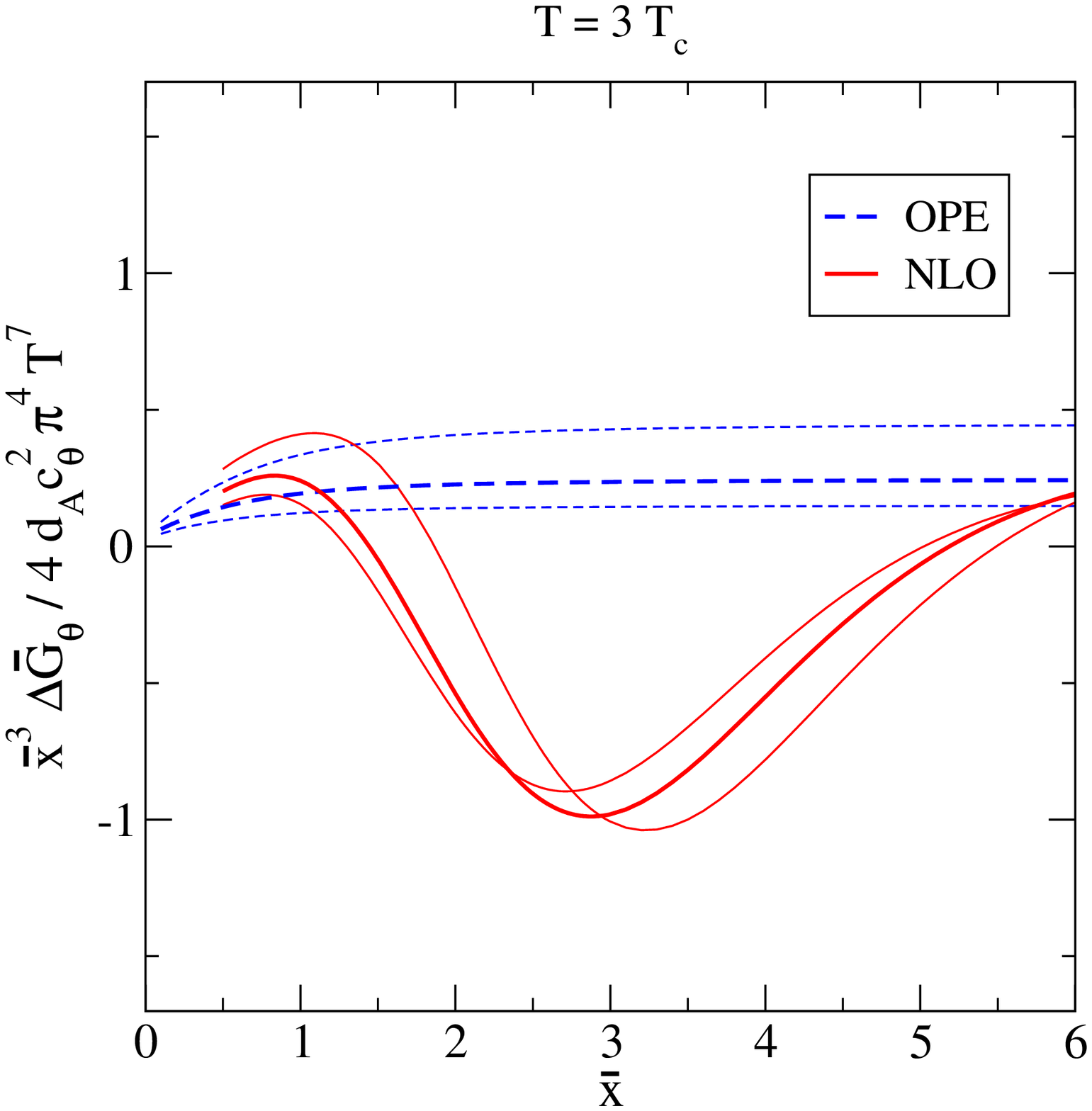}\hspace{0.4cm}\includegraphics[height=.3\textheight]{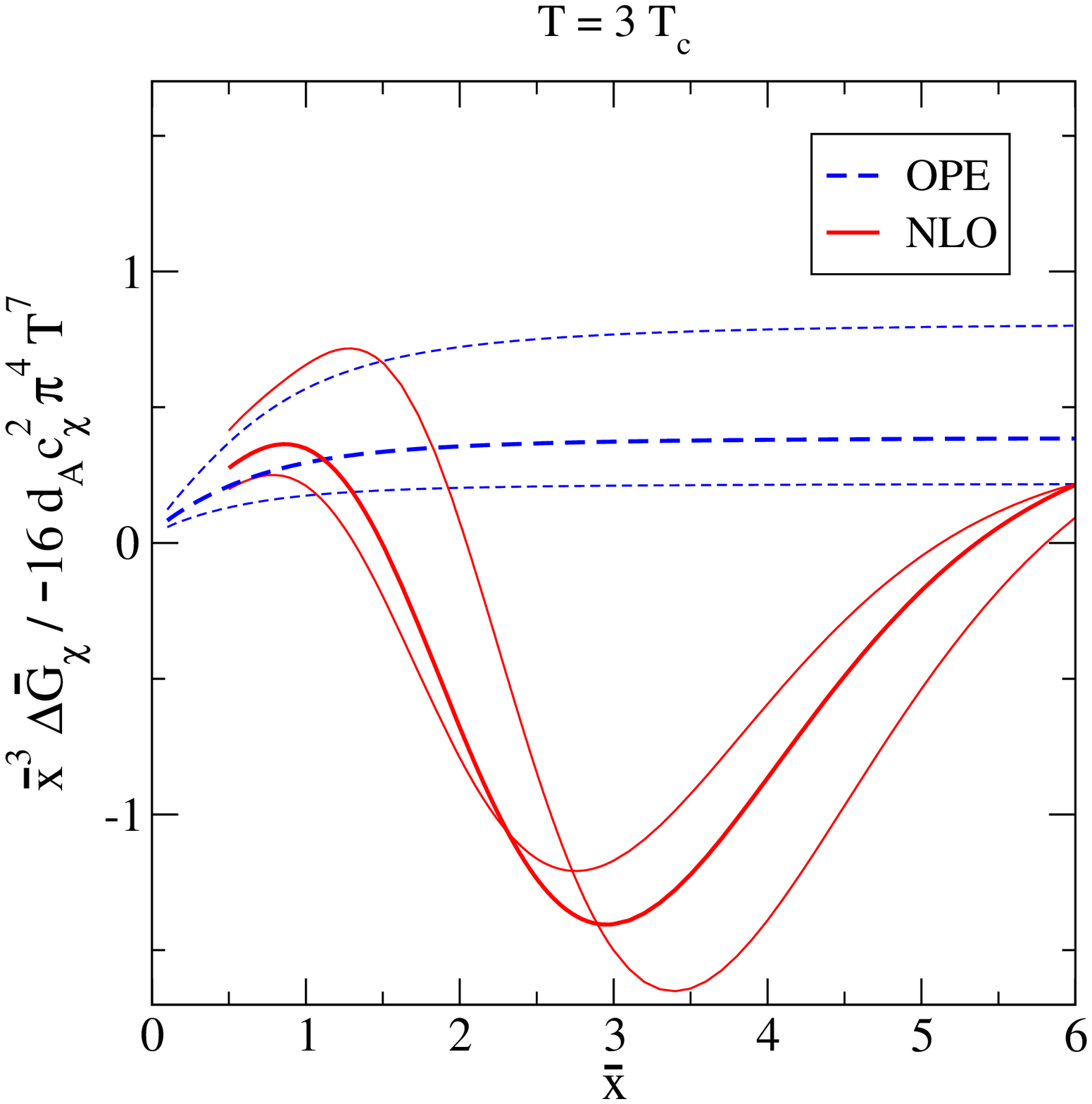}
\label{fig2}
\end{figure}

In Ref.~\cite{Laine:2010fe}, the considerably more daunting task of determining the full $r$-dependence of the spatial correlators --- this time averaged over the imaginary time $\tau$ --- was undertaken. The results in the two channels turned out to show remarkably similar behavior as well as quite satisfactory convergence properties, as demonstrated by Fig.~\ref{fig2}, where we display the vacuum subtracted correlators evaluated at $T=3T_c$ with $\bar{x} \equiv 2\pi T r$ ranging from 0 to 6. The quantities plotted are normalized in such a way that their free theory parts coincide, suggesting that the largest quantitative differences are obtained around $\bar{x}\approx 3$. Finally, to determine the correct large-distance behavior of the quantities, one needs to perform a leading order resummation calculation within the framework of the effective three-dimensional theory EQCD, which amounts to replacing the leading infrared behavior of the correlators ($\sim - g^6 \Nc/\bar{x}^4$) by an exponentially suppressed term, proportional to $\exp(-\mE r)$ in the pseudoscalar and $\exp(- 2 \mE r)$ in the scalar channel, with $\mE$ denoting the leading order electric screening mass. The transition from 'intermediate' to 'long' distances, signalled by the onset of exponential behavior in the correlators, is observed to take place at a distance scale $\bar{x}\sim 10$.

\section{Conclusions and Outlook}

In Refs.~\cite{lvv,Laine:2010fe}, we have investigated the behavior of the correlation functions of scalar and pseudoscalar operators in pure SU(3) Yang-Mills theory, determining in particular the full $r$-dependence of the time averaged Euclidean correlators. The results interpolate between the short distance regime, where they can be cast in the form of an OPE, and the infrared limit, where, as discussed in Ref.~\cite{Arnold:1995bh}, they reveal aspects of the screening properties of the medium. Ultimately, one would like to perform a detailed comparison of the present results with lattice data, once studies similar to those presented in Ref.~\cite{Iqbal:2009xz} emerge for the time averaged quantities. Already now, it can nevertheless be remarked that differences between the $\theta$ and $\chi$ channels are relatively subdued in the perturbative results, and it would most likely take at least a three-loop calculation to produce effects of a magnitude similar to those observed in Ref.~\cite{Iqbal:2009xz}.

Interestingly, the behavior of the time-averaged scalar and pseudoscalar correlators has very recently been determined in strongly coupled ${\mathcal N}=4$ SYM theory, where they are seen to coincide at least at the level of the supergravity approximation (leading order in $1/N$ and $1/\lambda$) \cite{kv}. The results thus obtained display a remarkable qualitative similarity to the perturbative ones, and it will be very interesting to see, how the lattice data will eventually compare with these two predictions.

Finally, there are a number of interesting directions, in which the present studies can be continued. Lattice determinations of transport coefficients, such as Ref.~\cite{Meyer:2007dy}, require as input detailed information on the behavior of the corresponding spectral density, which is a quantity that can be evaluated for general $\omega$ (\textit{i.e.~}not only in the UV limit) with methods very similar to those discussed in Refs.~\cite{lvv,Laine:2010fe}. A computationally considerably more demanding, but also highly interesting generalization would be to investigate the shear channel, paving the way to eventual improvements in the lattice determinations of the shear viscosity of Yang-Mills theory \cite{Meyer:2007ic}. Finally, adding fermions to the system both in the form of new operators and as 'sea quarks' is in principle straightforward, and should not lead to conceptual complications. These issues are currently under investigation.


\begin{theacknowledgments}

AV has been supported by the Sofja Kovalevskaja program of the Alexander von Humboldt foundation.

\end{theacknowledgments}

\end{document}